\documentclass[aps,prb,showpacs,twocolumn,superscriptaddress]{revtex4-1}
\usepackage{graphicx,amsmath,amssymb}
\usepackage[usenames]{color}
\usepackage{indentfirst}
\usepackage{float}
\usepackage[T1]{fontenc}
\usepackage[colorlinks=true, citecolor=blue, urlcolor=blue, linkcolor=blue ]{hyperref}
\hypersetup{breaklinks=true}
\begin{document}

\bibliographystyle{apsrev4-1}

\title{Solvable Periodic Anderson Model with Infinite-Range Hatsugai-Kohmoto Interaction: Ground-states and beyond}
\author{Yin Zhong}
\email{zhongy@lzu.edu.cn}
\affiliation{School of Physical Science and Technology $\&$ Key Laboratory for
Magnetism and Magnetic Materials of the MoE, Lanzhou University, Lanzhou 730000, China}
\affiliation{Lanzhou Center for Theoretical Physics, Key Laboratory of Theoretical Physics of Gansu Province}
\begin{abstract}
In this paper we introduce a solvable two-orbital/band model with infinite-range Hatsugai-Kohmoto interaction, which serves as a modified periodic Anderson model. Its solvability results from strict locality in momentum space, and is valid for arbitrary lattice geometry
and electron filling. Case study on a one-dimension ($1D$) chain shows that the ground-states have Luttinger theorem-violating non-Fermi-liquid-like metallic state, hybridization-driven insulator and interaction-driven featureless Mott insulator. The involved quantum phase transition between metallic and insulating states belongs to the universality of Lifshitz transition, i.e. change of topology of Fermi surface or band structure. Further investigation on $2D$ square lattice indicates its similarity with the $1D$ case, thus the findings in the latter may be generic for all spatial dimensions. We hope the present model or its modification may be useful for understanding novel quantum states in $f$-electron compounds, particularly the topological Kondo insulator candidate SmB$_{6}$ and YbB$_{12}$.
\end{abstract}

\maketitle
\section{Introduction}
Recently, solvable quantum many-body systems such as Sachdev-Ye-Kitaev, Kitaev's toric code and honeycomb model have attracted great interest due to emergent novel non-Fermi liquid and quantum spin liquid states.\cite{Sachdev,Maldacena,Chowdhury,Kitaev1,Kitaev2,Zhou,Prosko,Zhong2013,Smith2017,Ng2018} Among them, an infinite-range interaction model without any quenched disorder or local gauge structure called Hatsugai-Kohmoto (HK) model has been revisited.\cite{Hatsugai1992,Baskaran1991,Hatsugai1996,Phillips2018,Yeo2019,Phillips2020,Yang2021,Zhu2021,Zhao2022,Setty2021,Mai2022,Huang2022,Li2022} The original HK model provides a strictly exact example for non-Fermi liquid and featureless Mott insulator in any spatial dimension, which is rare in statistical mechanics and condensed matter physics. The solvability of HK model results from its locality in momentum space and one can diagonalize HK Hamiltonian (only $4\times4$-matrix) for each momentum. The current studies have mainly focused on an interesting extension of HK model, i.e. the superconducting instability from the intrinsic non-Fermi liquid state in HK model,\cite{Phillips2020} (note however a study on Kondo impurity in HK model\cite{Setty2021}) and unexpected properties (compared with standard Bardeen-Cooper-Schrieffer (BCS) model\cite{Bardeen1957}) have been discovered, e.g. the emergence of topological $s$-wave pairing, two-stage superconductivity, tricritical point and absence of Hebel-Slichter peak.\cite{Zhu2021,Zhao2022,Li2022}

In this paper, we introduce another extension of HK model, i.e. a two-orbital/band lattice electron system which can be considered as a modified periodic Anderson model (PAM),
\begin{eqnarray}
\hat{H}&=&-\sum_{i,j,\sigma}t_{ij}^{c}\hat{c}_{i\sigma}^{\dag}\hat{c}_{j\sigma}-\sum_{i,j,\sigma}t_{ij}^{f}\hat{f}_{i\sigma}^{\dag}\hat{f}_{j\sigma}+E_{f}\sum_{j\sigma}\hat{f}_{j\sigma}^{\dag}\hat{f}_{j\sigma}\nonumber\\
&+&V\sum_{j\sigma}(\hat{c}_{j\sigma}^{\dag}\hat{f}_{j\sigma}+\hat{f}_{j\sigma}^{\dag}\hat{c}_{j\sigma})-\mu\sum_{j\sigma}(\hat{c}_{j\sigma}^{\dag}\hat{c}_{j\sigma}+\hat{f}_{j\sigma}^{\dag}\hat{f}_{j\sigma})\nonumber\\
&+&\frac{U}{N_{s}}\sum_{j_{1},j_{2},j_{3},j_{4}}\delta_{j_{1}+j_{3}=j_{2}+j_{4}}
\hat{f}_{j_{1}\uparrow}^{\dag}\hat{f}_{j_{2}\uparrow}\hat{f}_{j_{3}\downarrow}^{\dag}\hat{f}_{j_{4}\downarrow}\label{eq1}.
\end{eqnarray}
Here, $\hat{c}_{j\sigma}^{\dag}$ is the creation operator of conduction electron ($c$-electron) while $\hat{f}_{j\sigma}^{\dag}$ denotes $f$-electron at site $j$. $t_{ij}^{c},t_{ij}^{f}$ are hopping integral between $i,j$ sites for $c$ and $f$-electron, respectively. Note that, $t_{ij}^{f}$ is zero in standard PAM and $f$-electron is strictly local in that case so one may call it local electron.\cite{Tsunetsugu} $E_{f}$ is the energy level of $f$-electron and the hybridization strength between $c$ and $f$-electron is $V$. (spin and site-dependent $V$ is also permitted and it leads to non-trivial quantum topological phase such as topological Kondo insulator,\cite{Dzero,Dzero2012,Griffith2019,Dzero2016,Ghazaryan2021} and Kondo liquid with hybridization node\cite{Ghaemi2007,Ghaemi2008}) Furthermore, chemical potential $\mu$ has been added to fix electron's density. $N_{s}$ is the number of sites. The last term of $\hat{H}$ is the less unfamiliar HK interaction,\cite{Hatsugai1992} (unlike the usual Hubbard interaction $U\sum_{j}\hat{f}_{j\uparrow}^{\dag}\hat{f}_{j\uparrow}\hat{f}_{j\downarrow}^{\dag}\hat{f}_{j\downarrow}$ in standard PAM) which is an infinite-range interaction between four electrons but preserves the center of motion for $f$-electron due to the constraint of $\delta$ function. This interaction plays a fundamental role in solving this model as we will see later.

Importantly, Eq.~\ref{eq1} is local in momentum space after Fourier transformation and the resultant Hamiltonian reads as $\hat{H}=\sum_{k}\hat{H}_{k}$,
\begin{eqnarray}
\hat{H}_{k}&=&\sum_{\sigma}(\varepsilon_{k}^{c}-\mu)\hat{c}_{k\sigma}^{\dag}\hat{c}_{k\sigma}+\sum_{\sigma}(\varepsilon_{k}^{f}+E_{f}-\mu)\hat{f}_{k\sigma}^{\dag}\hat{f}_{k\sigma}\nonumber\\
&+&V\sum_{\sigma}(\hat{c}_{k\sigma}^{\dag}\hat{f}_{k\sigma}+\hat{f}_{k\sigma}^{\dag}\hat{c}_{k\sigma})+U
\hat{f}_{k\uparrow}^{\dag}\hat{f}_{k\uparrow}\hat{f}_{k\downarrow}^{\dag}\hat{f}_{k\downarrow}\label{eq2},
\end{eqnarray}
where $\varepsilon_{k}^{c},\varepsilon_{k}^{f}$ are dispersion of electrons. It is emphasized that the locality of above Hamiltonian stems from infinite-range HK interaction preserving center of motion. In contrast, the Hubbard interaction in momentum space is rather nonlocal as $U\sum_{k,k',q}\hat{f}_{k+q\uparrow}^{\dag}\hat{f}_{k\uparrow}\hat{f}_{k'-q\downarrow}^{\dag}\hat{f}_{k'\downarrow}$, thus it cannot lead to simple formalism in our model.

Now, if we choose Fock state
\begin{equation}
|n_{1},n_{2},n_{3},n_{4}\rangle\equiv
(\hat{c}_{k\uparrow}^{\dag})^{n_{1}}|0\rangle(\hat{c}_{k\downarrow}^{\dag})^{n_{2}}|0\rangle(\hat{f}_{k\uparrow}^{\dag})^{n_{3}}|0\rangle
(\hat{f}_{k\downarrow}^{\dag})^{n_{4}}|0\rangle\label{eq3}
\end{equation}
with $n_{i}=0,1$ as basis, $\hat{H}_{k}$ can be written as a block-diagonal $16\times16$ matrix and a direct numerical diagonalization gives $16$ eigen-energy $E_{k}(i)$ and eigen-state $\psi_{k}(i)$.($i=1,2...16$ and $|\psi_{k}(1)\rangle$ is the ground-state for each $\hat{H}_{k}$). Details on $\hat{H}_{k}$'s matrix is shown in Appendix.A.

Therefore, the many-body ground-state of $\hat{H}$ is just the direct-product state of each $\hat{H}_{k}$'s ground-state, i.e. $|\Psi_{g}\rangle=\prod_{k}|\psi_{k}(1)\rangle$ with corresponding ground-state energy $E_{g}=\sum_{k}E_{k}(1)$. Similarly, excited states and their energy are easy to be constructed, so our model $\hat{H}$ (Eq.~\ref{eq1}) has been solved since all eigen-states and eigen-energy are found.

It is noted that the solvability of our model involves only locality in momentum space, therefore Eq.~\ref{eq1} is solvable for arbitrary lattice geometry, spatial dimension and electron filling, in contrast to standard PAM, where notorious fermion minus-sign problem and growth of quantum entanglement beyond area-law prevent exact solution or reliable numerical simulation.\cite{Vekic1995,Masuda2015,Schafer2019} Furthermore, our solvable model does not rely on disorder average and large-$N$ limit, which are crucial in classic Sherrington-Kirkpatrick's spin glass model and the more recent Sachdev-Ye-Kitaev model.\cite{Sherrington1975,Sachdev,Maldacena,Chowdhury} In addition, including pairing term like $\hat{c}_{k\uparrow}^{\dag}\hat{c}_{-k\downarrow}^{\dag}$,$\hat{f}_{k\uparrow}^{\dag}\hat{f}_{-k\downarrow}^{\dag}$ or $\hat{c}_{k\uparrow}^{\dag}\hat{f}_{-k\downarrow}^{\dag}$ as done in previous studies on superconductivity of HK,\cite{Phillips2020,Zhu2021,Zhao2022} does not change the solvability but only enlarges the dimension of Hamiltonian in momentum space.

The remaining part of this article is organized as follows. In Sec.~\ref{sec1}, we study a $1D$ chain from our model and its ground-state diagram has been established. Several physical quantities like particle density distribution, density of state, spectral function and Luttinger integral are calculated to characterize possible states. Sec.~\ref{sec2} is devoted to some discussions, i.e. the case study on $2D$ square lattice and the relation between particle density and metallic states. Brief summary is given in Sec.~\ref{sec3} and we also suggest that modification of our model may be relevant to understand the strong coupling physics in the topological Kondo insulator.
\section{An explicit example: the $1D$ model}\label{sec1}
\subsection{The ground-state}
Now, to extract the physics of our model, we focus on its $1D$ version, (extension to other lattice is straightforward) whose Hamiltonian reads as follows,
\begin{eqnarray}
\hat{H}&=&\sum_{k}\hat{H}_{k},\nonumber\\
\hat{H}_{k}&=&\sum_{\sigma}(\varepsilon_{k}-\mu)\hat{c}_{k\sigma}^{\dag}\hat{c}_{k\sigma}+\sum_{\sigma}(E_{f}-\mu)\hat{f}_{k\sigma}^{\dag}\hat{f}_{k\sigma}\nonumber\\
&+&V\sum_{\sigma}(\hat{c}_{k\sigma}^{\dag}\hat{f}_{k\sigma}+\hat{f}_{k\sigma}^{\dag}\hat{c}_{k\sigma})+U
\hat{f}_{k\uparrow}^{\dag}\hat{f}_{k\uparrow}\hat{f}_{k\downarrow}^{\dag}\hat{f}_{k\downarrow}\label{eq4},
\end{eqnarray}
where $\varepsilon_{k}=-2t\cos k$ is the $1D$ dispersion from nearest-neighbor-hopping $t$ and $f$-electron's dispersion is not included as in standard PAM. To be specific, we set $t=V=1$ as energy unit (smaller $V$ is more relevant to experiments in heavy fermion systems but physics will not be changed) and change $E_{f},\mu,U$ to explore the ground-state phase diagram.
\begin{figure}
\includegraphics[width=1.0\linewidth]{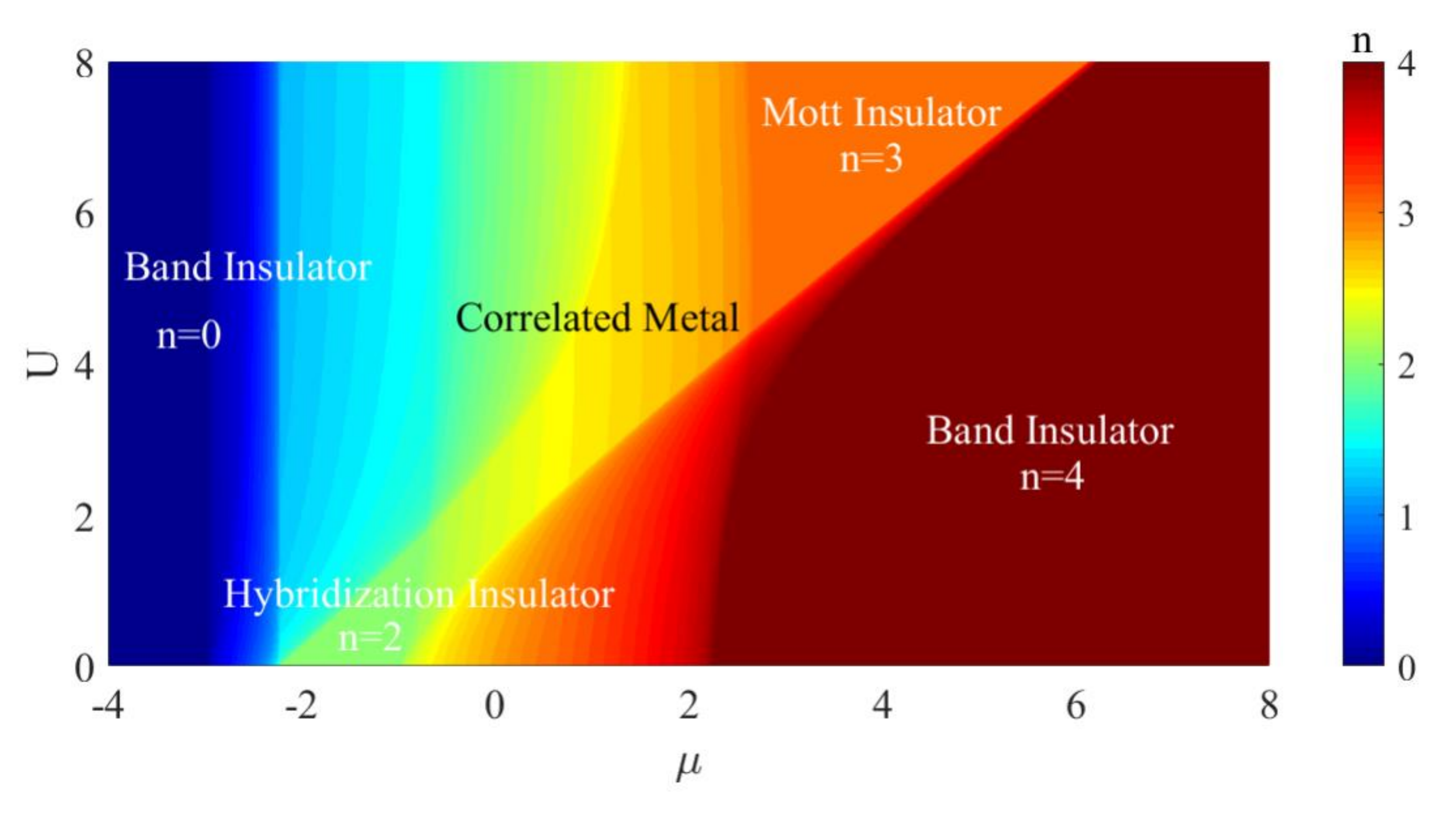}
\caption{\label{fig:1} Ground-state phase diagram of Eq.~\ref{eq4} with fixed $E_{f}=-2$. There exit three kinds of insulating states like band insulator, hybridization insulator, Mott insulator and one metallic state called correlated metal.}
\end{figure}

To characterize possible ground-state phases, we calculate some physical observable, e.g. the particle distribution function $n_{k}=\langle \hat{n}_{k}\rangle$,($\hat{n}_{k}=\hat{n}_{k}^{c}+\hat{n}_{k}^{f}=\sum_{\sigma}(\hat{c}_{k\sigma}^{\dag}\hat{c}_{k\sigma}+\hat{f}_{k\sigma}^{\dag}\hat{f}_{k\sigma}$) and $\langle...\rangle$ is chosen as average over $|\psi_{k}(1)\rangle$ if we focus on ground-state properties) density of state (DOS) of $c$-electron $N_{c}(\omega)$, $f$-electron $N_{f}(\omega)$ and the total one $N(\omega)$ and the spectral function of $c$-electron $A_{c}(k,\omega)$ and of $f$-electron $A_{f}(k,\omega)$.

To calculate the mentioned quantities, we first define retarded Green's function for $c$-electron in Heisenberg picture as $G_{\sigma}^{c}(k,t)=-i\theta(t)\langle [\hat{c}_{k\sigma}(t),\hat{c}_{k\sigma}^{\dag}]_{+}\rangle$. ($\theta(x)=1$ for $x>0$ and vanishes if $x<0$) Then, its Fourier transformation is denoted as $G_{\sigma}^{c}(k,\omega)$, which has the following Lehmann spectral representation,\cite{Coleman2015}
\begin{eqnarray}
G_{\sigma}^{c}(k,\omega)&=&\sum_{j=1}^{16}\frac{|\langle\psi_{k}(1)|\hat{c}_{k}|\psi_{k\sigma}(j)\rangle|^{2}}{\omega+i0^{+}+E_{k}(1)-E_{k}(j)}\nonumber\\
&+&\sum_{j=1}^{16}\frac{|\langle\psi_{k}(1)|\hat{c}_{k\sigma}^{\dag}|\psi_{k}(j)\rangle|^{2}}{\omega+i0^{+}+E_{k}(j)-E_{k}(1)}.\nonumber
\end{eqnarray}
At the same time, the retarded Green's function for $f$-electron has identical formalism with simple replacement $\hat{c}_{k\sigma}\rightarrow \hat{f}_{k\sigma}$. Therefore, $A_{c}(k,\omega)=\sum_{\sigma}A_{c}^{\sigma}(k,\omega)=\sum_{\sigma}-\frac{1}{\pi}\mathrm{Im}G_{\sigma}^{c}(k,\omega)$. As for DOS, we have the relation $N(\omega)=N_{c}(\omega)+N_{f}(\omega)=\frac{1}{N_{s}}\sum_{k}(A_{c}(k,\omega)+A_{f}(k,\omega))$.

In Fig.~\ref{fig:1}, we fix $E_{f}=-2$ and establish a ground-state phase diagram for different $\mu$ and $U$. (Other choice of $E_{f}$ is explored in Appendix.B and no physics is changed)

Here, particle density $n=0,4$ ($n=\frac{1}{N_{s}}\sum_{k}n_{k}$) correspond to trivial band insulator with empty or full occupation for each $k$-state. (the corresponding wave-functions are $\prod_{k}|0000\rangle_{k},\prod_{k}|1111\rangle_{k}$ if one utilizes the Fock state Eq.~\ref{eq3})

For $n=2$, we observe an insulator dominated by hybridization strength $V$ and we call it hybridization insulator (HI). Physically, the origin of HI can be understood from $U=0$ limit, and one has the following quasi-particle Hamiltonian
\begin{equation}
\hat{H}_{k}=\sum_{\sigma}(E_{k+}\hat{\alpha}^{\dag}_{k\sigma}\hat{\alpha}_{k\sigma}+E_{k-}\hat{\beta}^{\dag}_{k\sigma}\hat{\beta}_{k\sigma})\label{eq5}
\end{equation}
with the help of Bogoliubov transformation $\hat{\alpha}_{k\sigma}=\mu_{k}\hat{c}_{k\sigma}+\nu_{k}\hat{f}_{k\sigma}$ and $\hat{\beta}_{k\sigma}=-\nu_{k}\hat{c}_{k\sigma}+\mu_{k}\hat{f}_{k\sigma}$. The quasi-particle energy $E_{k\pm}=\frac{1}{2}(\varepsilon_{k}+E_{f}\pm\sqrt{(\varepsilon_{k}-E_{f})^{2}+4V^{2}})-\mu$ and $\mu_{k}^{2}=\frac{1}{2}\left(1+\frac{\varepsilon_{k}-E_{f}}{\sqrt{(\varepsilon_{k}-E_{f})^{2}+4V^{2}}}\right)=1-\nu_{k}^{2}$. Then, if the lower band $E_{k-}$ is fully occupied, but the upper band $E_{k+}$ is empty, an insulator (with ground-state wave-function $\prod_{k\sigma}\hat{\beta}^{\dag}_{k\sigma}|0\rangle$) with particle density $n=2$ appears, which has the indirect gap $\sim V$ and direct gap $\Delta\equiv\mathrm{min}(E_{k+})-\mathrm{max}(E_{k+})\sim \frac{V^{2}}{t}$.

Next, let us examine the effect of finite $U$. When $U<\Delta$, we expect the gap does not close and the adiabatic principle of Landau applies, thus the system is still in HI but with renormalized gap and dispersion. Using parameters in our model ($t=V=1$), we find $\Delta\sim1$ and HI is stable if $U<\Delta\sim1$. This is indeed the case in Fig.~\ref{fig:1} and we think the above picture is justified. In addition, we note that in standard PAM, HI evolves into the antiferromagnetic insulator (on square lattice) or Kondo insulator when Hubbard-$U$ increases from $U=0$.\cite{Vekic1995}

\begin{figure}
\includegraphics[width=1.0\linewidth]{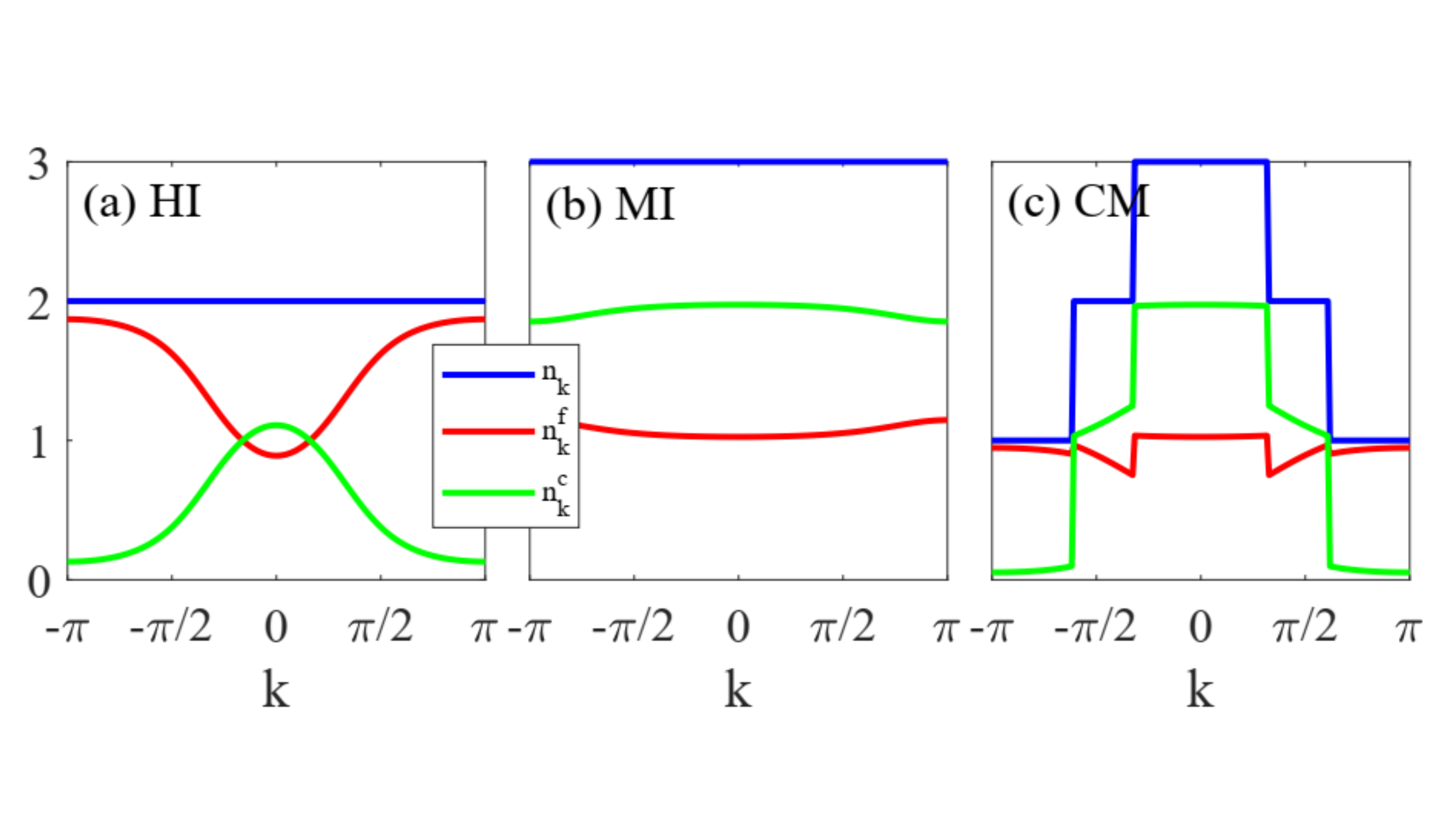}
\caption{\label{fig:2} Particle density distribution $n_{k},n_{k}^{f}$ and $n_{k}^{c}$ versus momentum $k$ with $E_{f}=-2$. (a) HI with $U=0.5,\mu=-1.3$; (b) MI with $U=6,\mu=3$; (c) CM with $U=6,\mu=0$.}
\end{figure}
\begin{figure}
\includegraphics[width=1.05\linewidth]{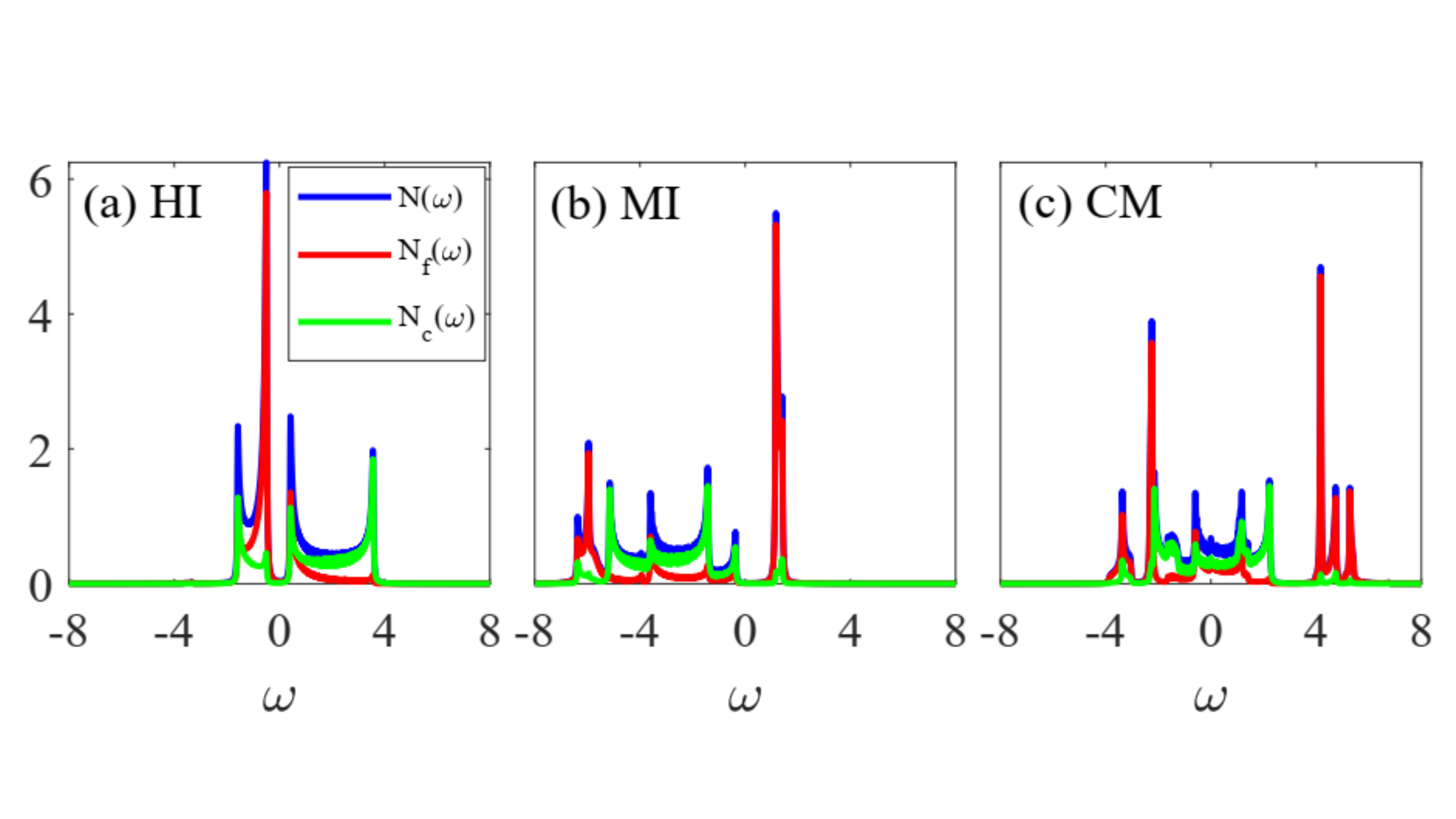}
\caption{\label{fig:3} Density of state $N(\omega),N_{f}(\omega)$ and $N_{c}(\omega)$ with $E_{f}=-2$. (a) HI with $U=0.5,\mu=-1.3$; (b) MI with $U=6,\mu=3$; (c) CM with $U=6,\mu=0$.}
\end{figure}
Next, we turn to $n=3$ regime, which is clearly driven by interaction and it can be denoted as Mott insulator (MI). From Fig.~\ref{fig:2}(b), we see that in MI, the particle density distribution $n_{k}$ is fixed to $3$ for each momentum, however, neither $n_{k}^{c}$ nor $n_{k}^{f}$ is fixed to integer, (so this state cannot be an orbital-selective Mott insulator by definition\cite{Pepin2007}) which should be compared with the ones in HI (Fig.~\ref{fig:2}(a)).

\begin{figure}
\includegraphics[width=1.2\linewidth]{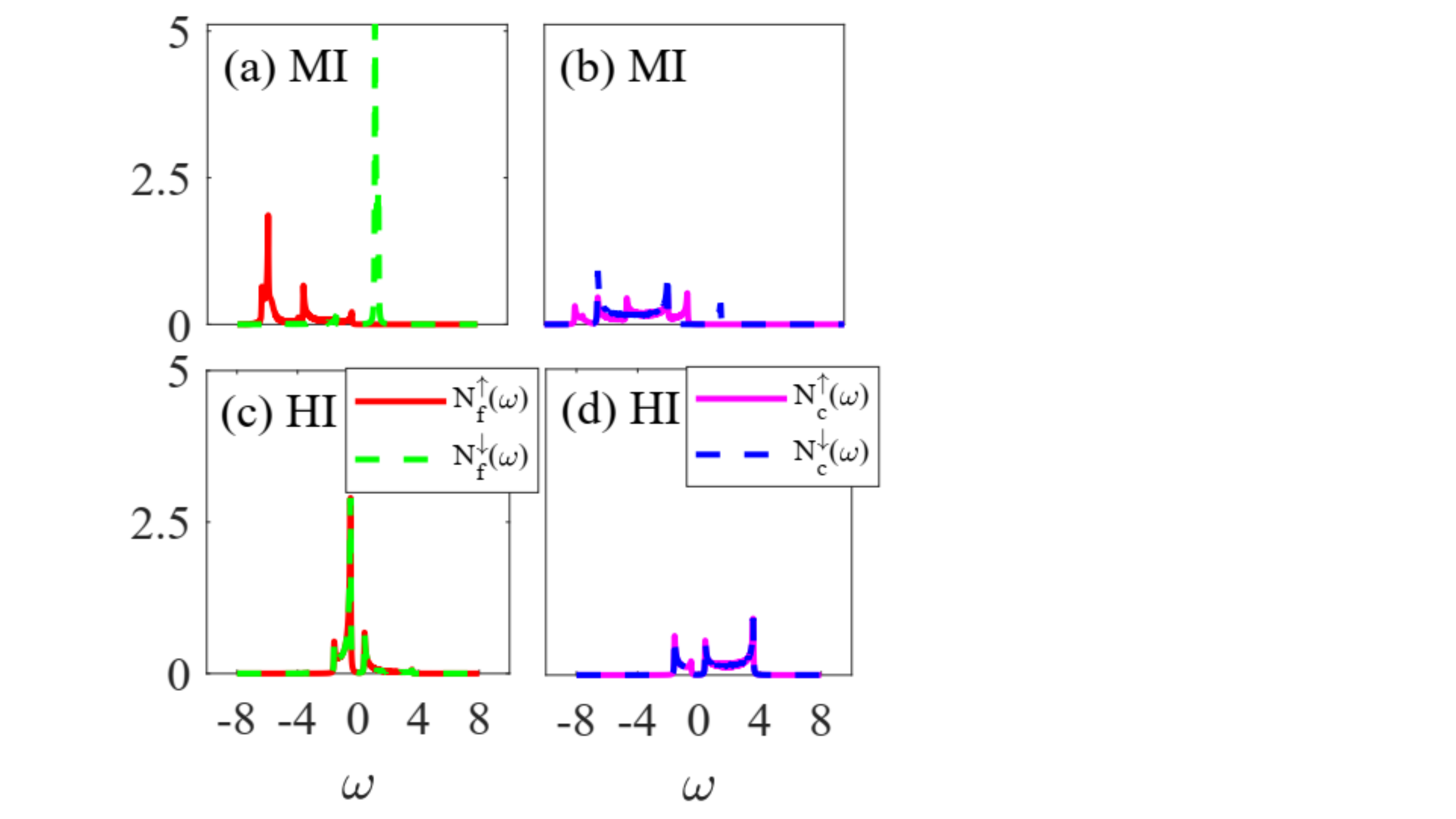}
\caption{\label{fig:7} The spin-resolved density of state for $f$ and $c$-electron $N_{f}^{\uparrow}(\omega),N_{f}^{\downarrow}(\omega),N_{c}^{\uparrow}(\omega),N_{c}^{\downarrow}(\omega)$ with $E_{f}=-2$. (a)(b) MI with $U=6,\mu=3$; (c)(d) HI with $U=0.5,\mu=-1.3$.}
\end{figure}

Furthermore, Fig.~\ref{fig:3} shows the DOS in HI (Fig.~\ref{fig:3}(a)) and MI (Fig.~\ref{fig:3}(b)). Although both MI and HI have noticeable gap near $\omega=0$, it is clear that a larger gap opens in MI than in HI and the former exhibits strong asymmetry for its DOS. Actually, a closer look at MI's spin-resolved DOS in Fig.~\ref{fig:7}(a)(b) implies that the DOS above Fermi energy ($\omega=0$) is dominated by spin-down electron while the $\omega<0$ part is contributed mainly from spin-up electron. This property seems to be the generic feature for HK-like models. Instead, as can seen in Fig.~\ref{fig:7}(c)(d), the spin degree of freedoms are degenerated in HI, (see e.g. the quasi-particle Hamiltonian Eq.~\label{eq5}) and they contribute equally to DOS ($N_{f}^{\uparrow}=N_{f}^{\downarrow},N_{c}^{\uparrow}=N_{c}^{\downarrow}$ in HI).

\begin{figure}
\includegraphics[width=1.0\linewidth]{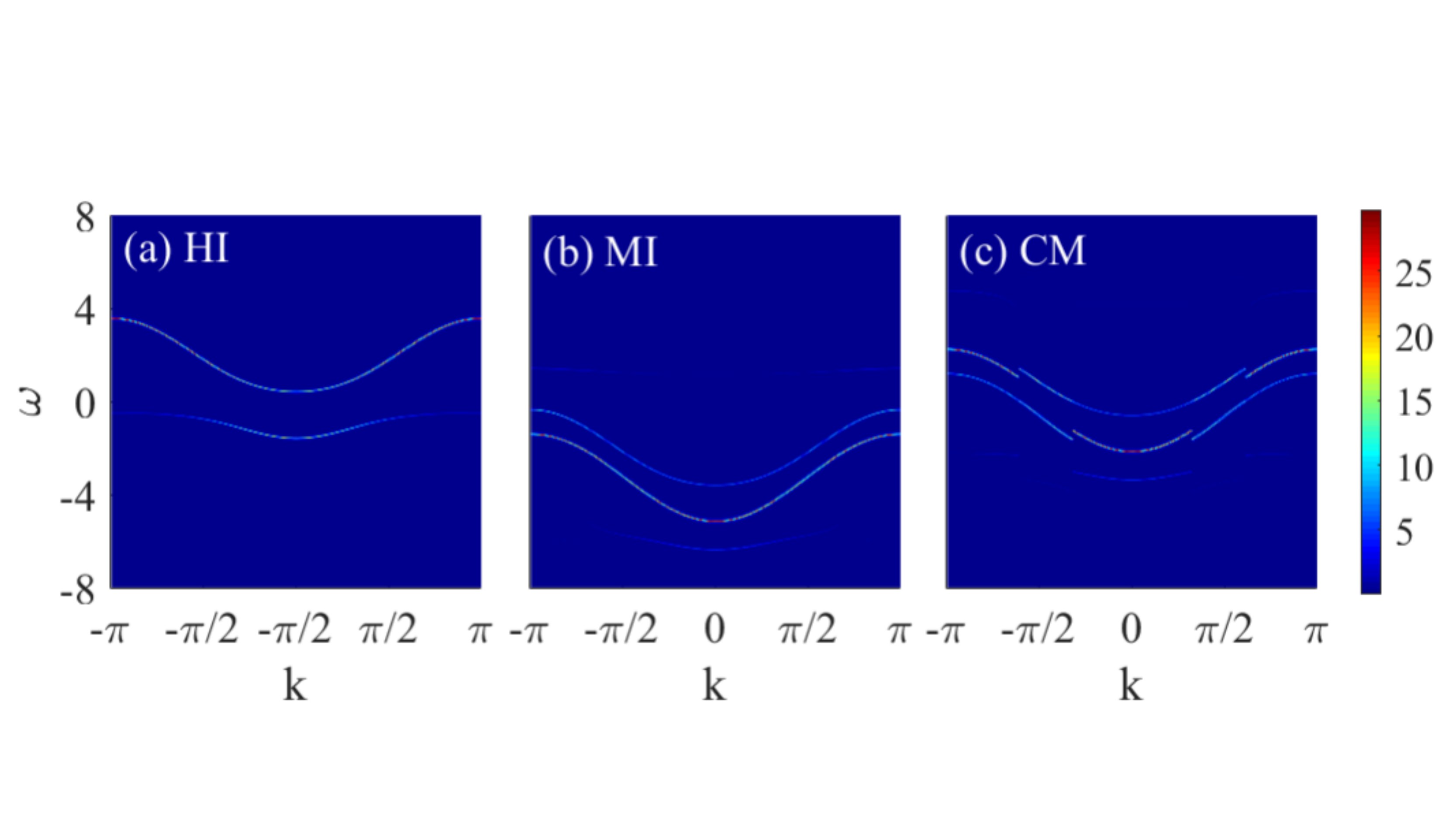}
\caption{\label{fig:4} Spectral function of $c$-electron $A_{c}(k,\omega)$ with $E_{f}=-2$. (a) HI with $U=0.5,\mu=-1.3$; (b) MI with $U=6,\mu=3$; (c) CM with $U=6,\mu=0$.}
\end{figure}

\begin{figure}
\includegraphics[width=1.0\linewidth]{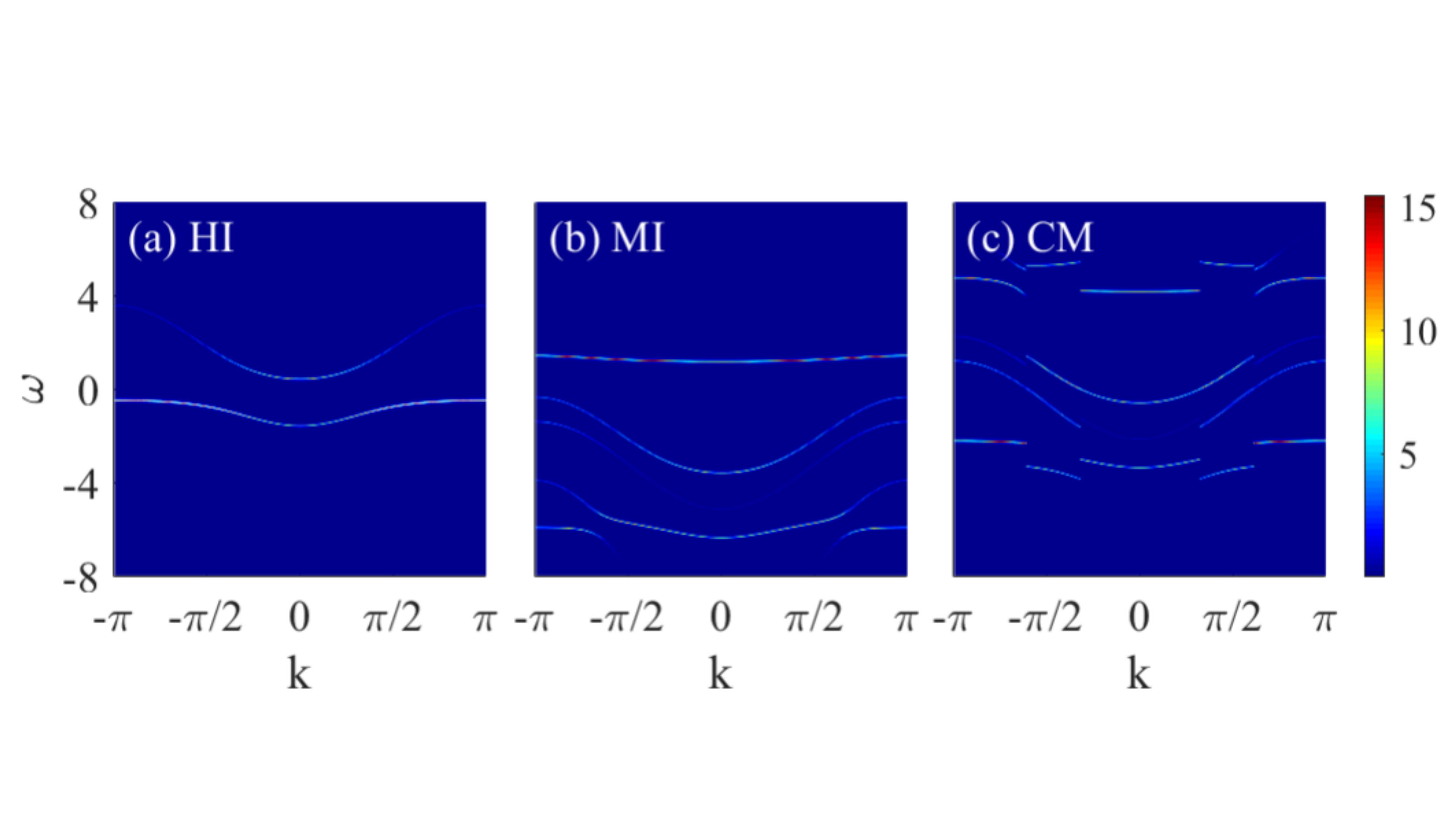}
\caption{\label{fig:5} Spectral function of $f$-electron $A_{f}(k,\omega)$ with $E_{f}=-2$. (a) HI with $U=0.5,\mu=-1.3$; (b) MI with $U=6,\mu=3$; (c) CM with $U=6,\mu=0$.}
\end{figure}

The spectral function of $c$ and $f$-electron $A_{c/f}(k,\omega)=\sum_{\sigma}A_{c/f}^{\sigma}(k,\omega)$ is plotted in Fig.~\ref{fig:4} and Fig.~\ref{fig:5}. We have checked that the spin-resolved spectral function $A_{c}^{\sigma},A_{f}^{\sigma}$ satisfy the sum-rule $\int_{-\infty}^{\infty} d\omega A_{c/f}^{\sigma}(k,\omega)=1$. It is observed that in MI, (Fig.~\ref{fig:4}(b) and Fig.~\ref{fig:5}(b)) the $c$-electron has dispersive band below Fermi energy ($\omega=0$) while a rather flat band appears above Fermi energy for $f$-electron. In contrast, two dispersive bands exist near Fermi energy for both of $c$ and $f$-electron in HI, (Fig.~\ref{fig:4}(a) and Fig.~\ref{fig:5}(a)) which embody the well-defined quasi-particle band $E_{k\pm}$.

Finally, the remaining state in the phase diagram is just the metallic one, which is denoted as correlated metal (CM). The typical particle density distribution of CM has been shown in Fig.~\ref{fig:2}(c). It is found that, in contrast with Fermi Liquid (FL) or more simply the Fermi gas, $n_{k}$ (also for $n_{k}^{c},n_{k}^{f}$) has two-jump behavior at certain momentum (one may call it Fermi point), whose structure is comparable with the non-Fermi liquid state in HK model.\cite{Hatsugai1992} Therefore, it indicates CM should not be a FL but non-Fermi liquid. The corresponding DOS and spectral function of CM have been investigated in Fig.~\ref{fig:3}(c), Fig.~\ref{fig:4}(c) and Fig.~\ref{fig:5}(c), and an interesting multi-band structure is visible though we have not found any analytical expression to explain this structure. (According to Appendix.A, we have $11$ different energy level for each $k$, thus there exist at most $11$ band-like structure in DOS or spectral function.)

\begin{figure}
\includegraphics[width=1.0\linewidth]{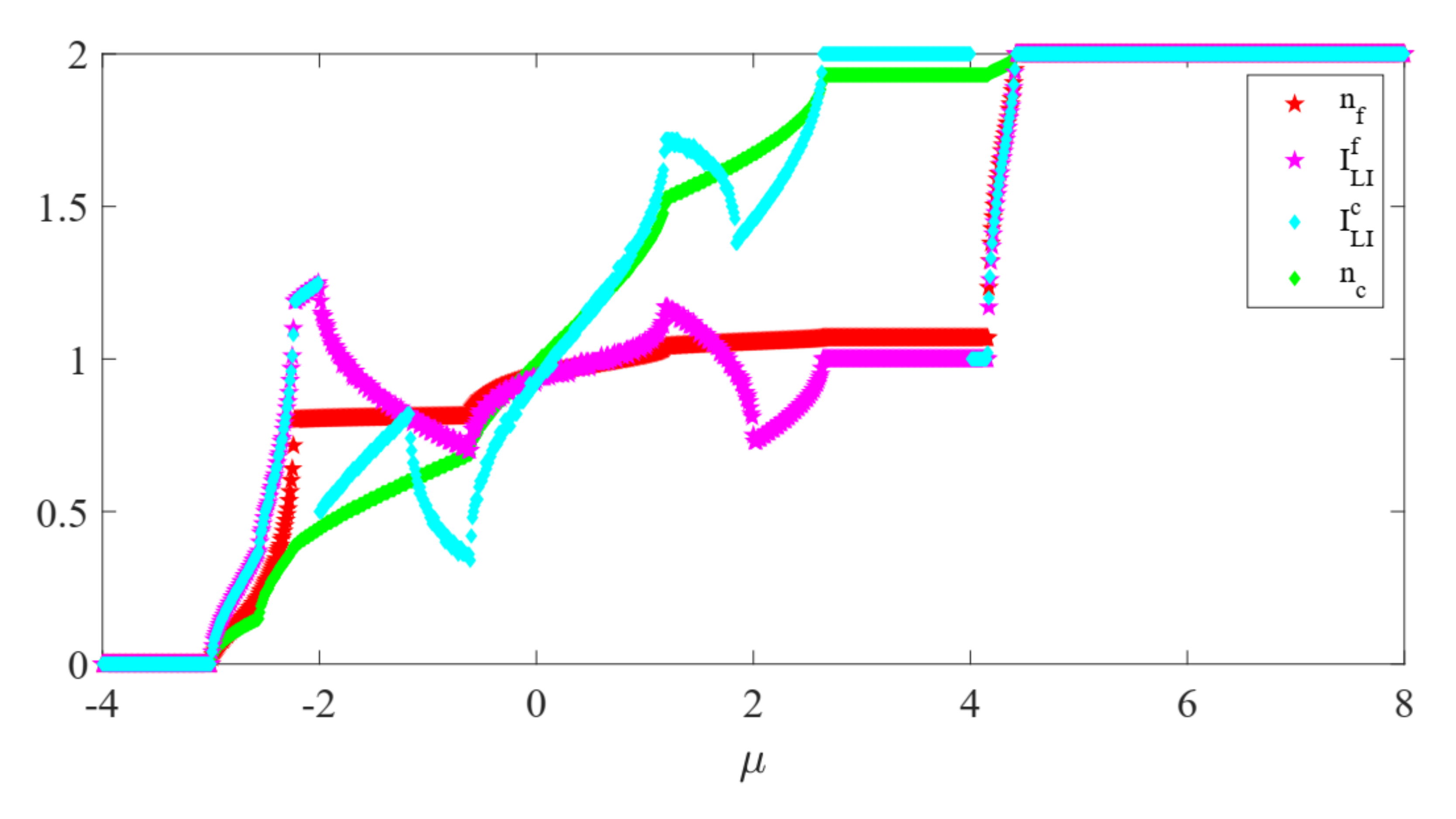}
\caption{\label{fig:6} Luttinger integral ($I_{LI}^{c},I_{LI}^{f}$) versus particle density $(n_{c},n_{f})$ for different chemical potential $\mu$ with $E_{f}=-2,U=6$.}
\end{figure}

To encode the nature of CM, we examine whether CM satisfies the Luttinger theorem,\cite{Luttinger1960a,Luttinger1960b,Oshikawa2000,Else2021} which has been believed to be defining feature of FL. The Luttinger theorem tells us that for FL-like states, (including Luttinger liquid in $1D$) the following integral, i.e. Luttinger integral, must be equal to particle density,\cite{Dzyaloshinskii2003}(here we use its $1D$ version)
\begin{equation}
I_{LI}=\sum_{\sigma}\int \frac{dk}{2\pi} \theta(\mathrm{Re}G_{\sigma}(k,\omega=0)).
\end{equation}
In above formula, the $\theta$-function counts the positive part of the real part of the retarded Green's function at zero frequency.
In our case, we have two Luttinger integral $I_{LI}^{c},I_{LI}^{f}$ for $c$ and $f$-electron. In Fig.~\ref{fig:6}, we have plotted $I_{LI}^{c},I_{LI}^{f}$ versus particle density $(n_{c},n_{f})$ for different chemical potential $\mu$ with fixed $E_{f}=-2,U=6$.
It is obvious that $I_{LI}^{c}\neq n_{c}$ and $I_{LI}^{f}\neq n_{f}$, thus the CM state ($-3<\mu<4$) violates the Luttinger theorem and it is indeed a non-Fermi liquid.

A careful reader may notice that $I_{LI}^{c},I_{LI}^{f}$ are nonmonotonic when $\mu$ changes. A straightforward explanation seems to be that such behavior results from the change of topology of Fermi surface or band structure, i.e. the famous Lifshitz transition. When Lifshitz transition appears with tuning $\mu$, the particle density will show kink-like structure. In Fig.~\ref{fig:6}, one see that when kinks emerge in $n_{c},n_{f}$, the Luttinger integral $I_{LI}^{c},I_{LI}^{f}$ exhibit strong deviation from Luttinger theorem. Therefore, CM is not a single phase but including many Lifshitz transitions.
\subsection{How about the phase transition?}
We have explored the ground-state phase diagram and there exit four states namely BI, HI, MI and CM. Except for the trivial BI, we expect quantum transitions between metallic and insulating states (CM-MI and CM-HI). Since there is no explicit candidate Landau's order parameter with noticeable symmetry-breaking, a natural guess suggests these transitions are of nature of the Lifshitz transition. One may argue that an alternative explanation can be certain topological order,\cite{Wen1990,Senthil2001,Oshikawa2006} but such possibility in $1D$ is not plausible since protypical $Z_{2}$ topological order is not stable in $1D$ unless one ignores the effect of effective electric field in $Z_{2}$ lattice gauge field theory.\cite{Kogut1979,Prosko} The possibility of topological order in $2D$ version of our model or other HK-like models have not been explored and we suspect that it will not relate to these mentioned models. After all, in this work, we only examine the Lifshitz transition.

To locate and characterize such putative transition, we inspect the behavior of charge susceptibility $\chi_{c}=\frac{\partial n}{\partial \mu}$, which diverges at critical point driven by charge fluctuation like Lifshitz transition.
\begin{figure}
\includegraphics[width=1.05\linewidth]{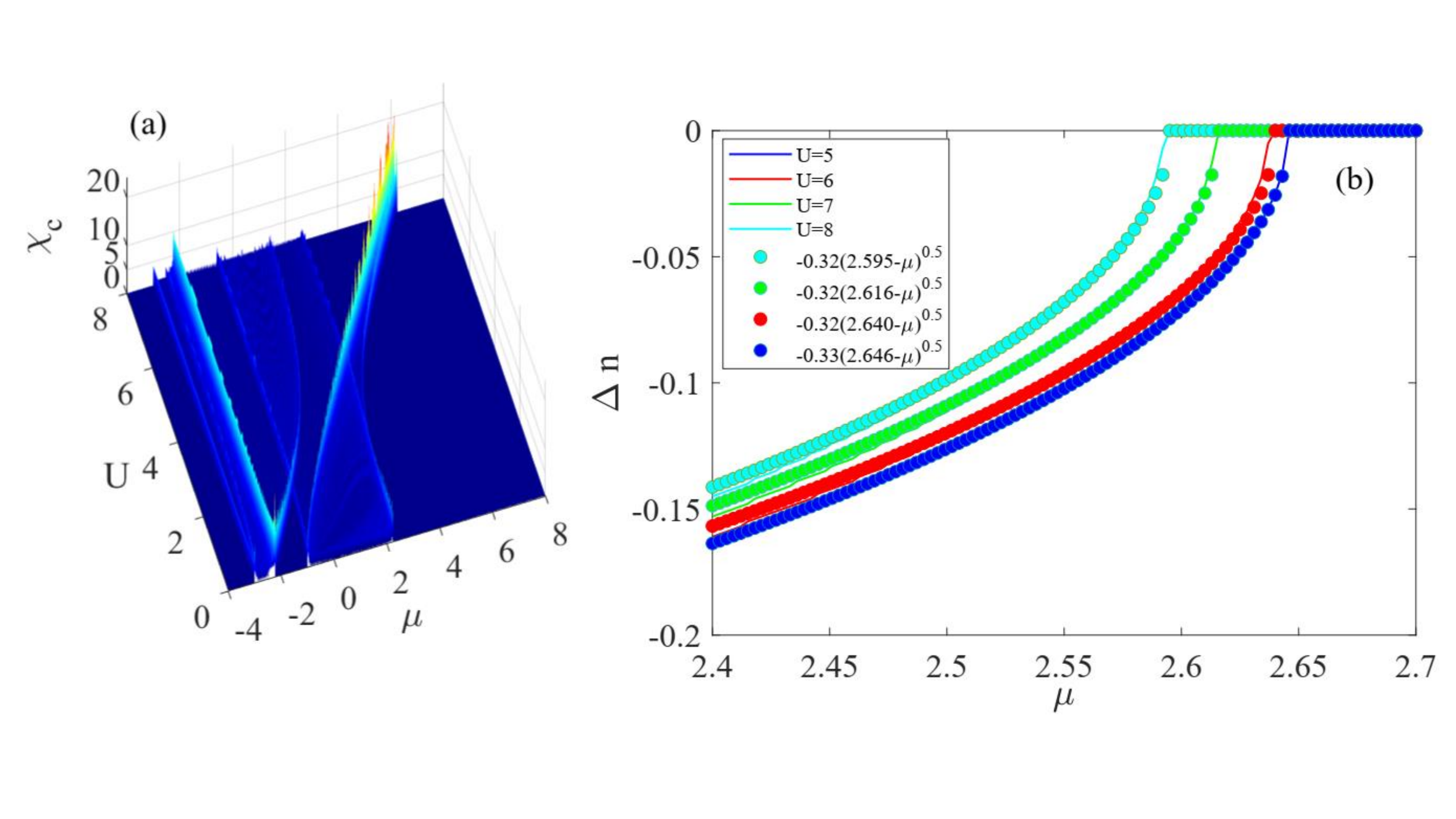}
\caption{\label{fig:8} (a) The charge susceptibility $\chi_{c}$ with $E_{f}=-2$; (b) Particle density difference $\Delta n$ near critical point versus fitted scaling formula Eq.~\ref{eq7} with $E_{f}=-2$.}
\end{figure}

As what can be seen in Fig.~\ref{fig:8}(a), the divergent $\chi_{c}$ is able to locate phase boundary, which agrees perfectly with our previous phase diagram Fig.~\ref{fig:1}.

Now, let we focus on the phase transition between CM and MI. To be specific, consider $U=5,6,7,8$ while tuning $\mu$ with fixed $E_{f}=-2$. These are the chemical potential-driven transitions and their critical points locate in $\mu=\mu_{c}=2.595,2.616,2.640,2.646$. According to general ideas of quantum phase transition, we expect the particle density near critical point has the following scaling form,\cite{Continentino}
\begin{equation}
\Delta n=n-n_{0}\sim (\mu-\mu_{c})^{\beta},\label{eq7}
\end{equation}
where $n_{0}$ denotes certain background which should be subtracted, $\mu_{c}$ is the location of critical point and $\beta$ is the critical exponent. The above scaling formula works well near critical points as shown in Fig.~\ref{fig:8}(b), and one finds that $\beta\simeq0.5$, which is the critical exponent for expected Lifshitz transition. ($\beta=d/2$ for free fermion Lifshitz transition universality with dimension of space $d$ and the dynamic critical exponent $z=2$) Furthermore, the ground-state energy and charge susceptibility also have scaling form $E_{g}-E_{g}^{0}\sim (\mu-\mu_{c})^{(d+2)/2}$, $\chi_{c}\sim (\mu-\mu_{c})^{(d-2)/2}$,\cite{Continentino,Vitoriano2000} which are consistent with our calculation if choosing spatial dimension $d=1$. Therefore, we conclude that the chemical potential-driven CM-MI transition is in fact the Lifshitz transition.

One may note that there also exists interaction-driven CM-MI transition. Although we have not explored its properties in detail, based on our knowledge on HK-like models, this transition should belong to the universality of Lifshitz transition, as well.\cite{Continentino,Vitoriano2000} In addition, it is not surprise to find that the CM-HI transition is of the nature of Lifshitz transition and we will not investigate it further.

\subsection{Is there any effective field theory description for CM, MI or HI?}
In the study of correlated electron systems, the effective (field) theory description is very useful since it can give us the low-energy theory, which is truly responsible for understanding on low-temperature/low-frequency thermodynamics and transport. Frankly speaking, we have no idea to construct a suitable effective theory for CM and MI. At this point, one may ask why the state-of-art $1D$ bosonization is not able to attack our model. The reason is that the Luttinger theorem, which fixes the Fermi point (Fermi surface in $1D$) for given particle density,\cite{Giamarchi} is violated in our model, thus the starting point of bosonization, i.e. the low-energy expansion around Fermi point, is meaningless. Consequently, we do not expect bosonization technique is useful to our model. (We have also tried to relate metallic CM to conformal-field-theory (CFT) but have failed since the finite-size scaling of CM in our model is not consistent with well-known CFT models, like minimal models.\cite{Francesco})

Next, one may note that our model is similar to the original HK model, and the latter one has many simple and beautiful analytic results. Thus, if we can project our model into HK model, life will be easy. Let us try this idea in terms of path integral formalism. The imaginary-time action of our model is just like
\begin{eqnarray}
S&=&\sum_{k}\int d\tau \sum_{\sigma}\bar{c}_{k\sigma}(\partial_{\tau}+\varepsilon_{k}-\mu)c_{k\sigma}+\sum_{\sigma}\bar{f}_{k\sigma}(\partial_{\tau}+E_{f}-\mu)f_{k\sigma}
\nonumber\\&+&V\sum_{\sigma}(\bar{c}_{k\sigma}f_{k\sigma}+\bar{f}_{k\sigma}c_{k\sigma})+U\bar{f}_{k\uparrow}f_{k\uparrow}\bar{f}_{k\downarrow}f_{k\downarrow},\label{eq9}
\end{eqnarray}
where $\bar{c}_{k\sigma},c_{k\sigma},\bar{f}_{k\sigma},f_{k\sigma}$ are anti-commutative Grassman field. Note that the HK interaction is active for $f$-electron, we may integrate $c$-electron out to get a $f$-electron-only theory. After integrating out c-electron's degree of freedom, we find the following action for $f$-electron,
\begin{eqnarray}
S_{f}&=&\sum_{k}\int d\tau\sum_{\sigma}\bar{f}_{k\sigma}(\partial_{\tau}+E_{f}-\mu)f_{k\sigma}+U\int d\tau\bar{f}_{k\uparrow}f_{k\uparrow}\bar{f}_{k\downarrow}f_{k\downarrow}\nonumber\\
&+&\int d\tau\int d\tau'\sum_{\sigma}\bar{f}_{k\sigma}(\tau)V^{2}G_{k}^{0}(\tau-\tau')f_{k\sigma}(\tau').
\end{eqnarray}
Here, $G_{k}^{0}(\tau)$ is the Fourier transformation of free $c$-electron Green's function $G^{0}(k,\omega_{n})=(i\omega_{n}-\varepsilon_{k}+\mu)^{-1}$. We see that the above action is nonlocal in imaginary-time thus cannot be written as the HK model.

As for HI, in the previous section, we have analyzed its $U=0$ limit and argued that it is stable if the gap still opens. Here, motivated by Eq.~\ref{eq9}, we treat HK interaction as perturbation and the induced correction at first order gives rise to the Hartree self-energy $\Sigma_{\sigma}(k,\omega)=Un_{k\bar{\sigma}}^{f}$.($n_{k\bar{\sigma}}^{f}=\langle \hat{f}_{k\bar{\sigma}}^{\dag}\hat{f}_{k\bar{\sigma}}\rangle$ with $\bar{\sigma}\equiv-\sigma$) Therefore, we have $f$-electron Green's function
\begin{equation}
G_{\sigma}^{f}(k,\omega)=\frac{1}{(G_{0}^{f}(k,\omega))^{-1}-Un_{k\bar{\sigma}}^{f}}
\end{equation}
with $G_{0}^{f}(k,\omega)=(\omega-E_{f}+\mu-V^{2}(\omega-\varepsilon_{k}+\mu)^{-1})^{-1}$ being the $f$-electron Green's function in $U=0$ limit. Now, the pole of $G_{\sigma}^{f}(k,\omega)$ determines the renormalized quasi-particle dispersion, whose form is
\begin{eqnarray}
&&\tilde{E}_{k\sigma\pm}=\frac{1}{2}[\varepsilon_{k}+E_{f}+Un_{k\bar{\sigma}}^{f}\pm A_{k\sigma}]-\mu\nonumber\\
&&A_{k\sigma}=\sqrt{(\varepsilon_{k}-E_{f})^{2}+(Un_{k\bar{\sigma}}^{f})^{2}+2Un_{k\bar{\sigma}}^{f}(3\varepsilon_{k}+E_{f}-4\mu)}.\nonumber
\end{eqnarray}

\section{Discussion}\label{sec2}
\subsection{Example on $2D$ square lattice}
In previous section, we have investigated a $1D$ model and established its phase diagram. In this subsection, we will briefly discuss the properties on the $2D$ square lattice. For this lattice, the only difference from $1D$ model is that the nearest-neighbor-hopping for $c$-electron generates $\varepsilon_{k}=-2t(\cos k_{x}+\cos k_{y})$ instead of $-2t\cos k$. Others are the same with our previous discussion.

\begin{figure}
\includegraphics[width=1.0\linewidth]{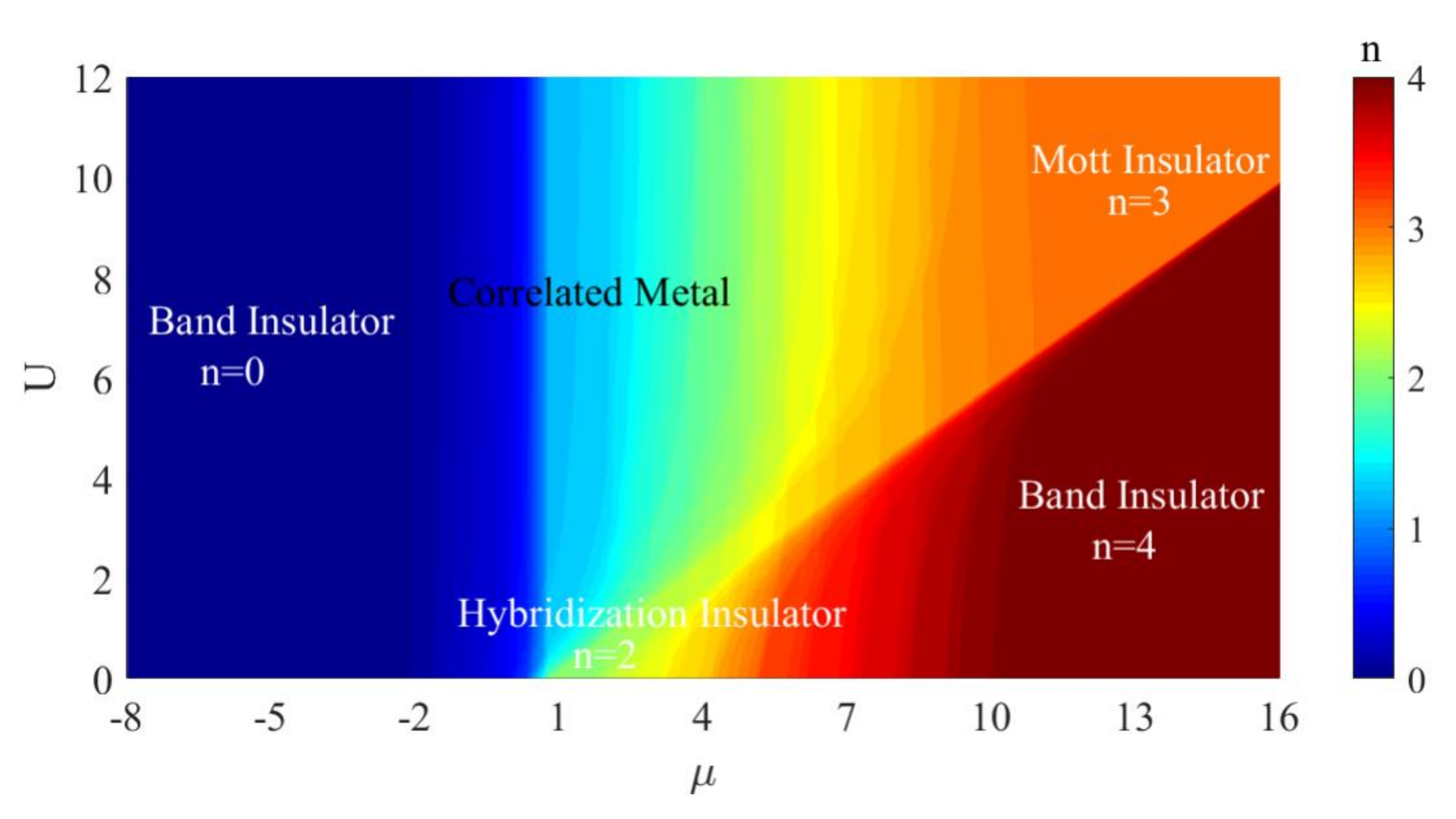}
\caption{\label{fig:9} Ground-state phase diagram of $2D$ square lattice with fixed $E_{f}=-2$. Note that it is similar with $1D$ case Fig.~\ref{fig:1}.}
\end{figure}

To get rough intuition on this $2D$ problem, we have plotted its ground-state phase diagram in Fig.~\ref{fig:9} with $t=V=1,E_{f}=-2$. It is seen that the structure of this phase diagram is quite similar to the $1D$ version Fig.~\ref{fig:1}, therefore we expect that the findings in the previous $1D$ model may be generic for all spatial dimensions.

\begin{figure}
\includegraphics[width=1.0\linewidth]{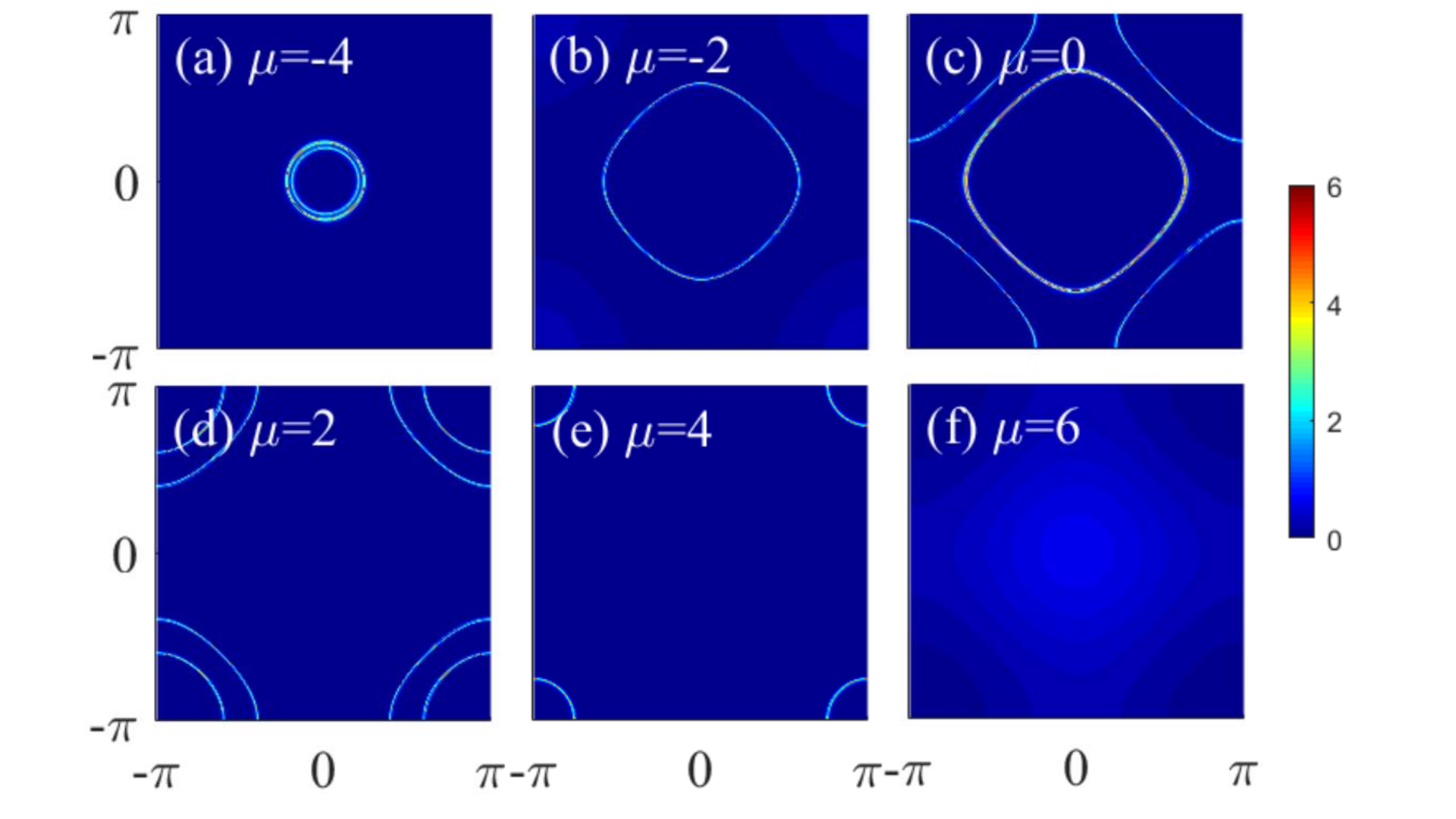}
\caption{\label{fig:10} The $f$-electron's zero-frequency spectral function $A_{f}(k_{x},k_{y},\omega=0)$ on $2D$ square lattice with $E_{f}=-2,U=8$. (a) $\mu=-4$; (b) $\mu=-2$; (c) $\mu=0$; (d) $\mu=2$; (e) $\mu=4$; (f) $\mu=6$.}
\end{figure}

\begin{figure}
\includegraphics[width=1.1\linewidth]{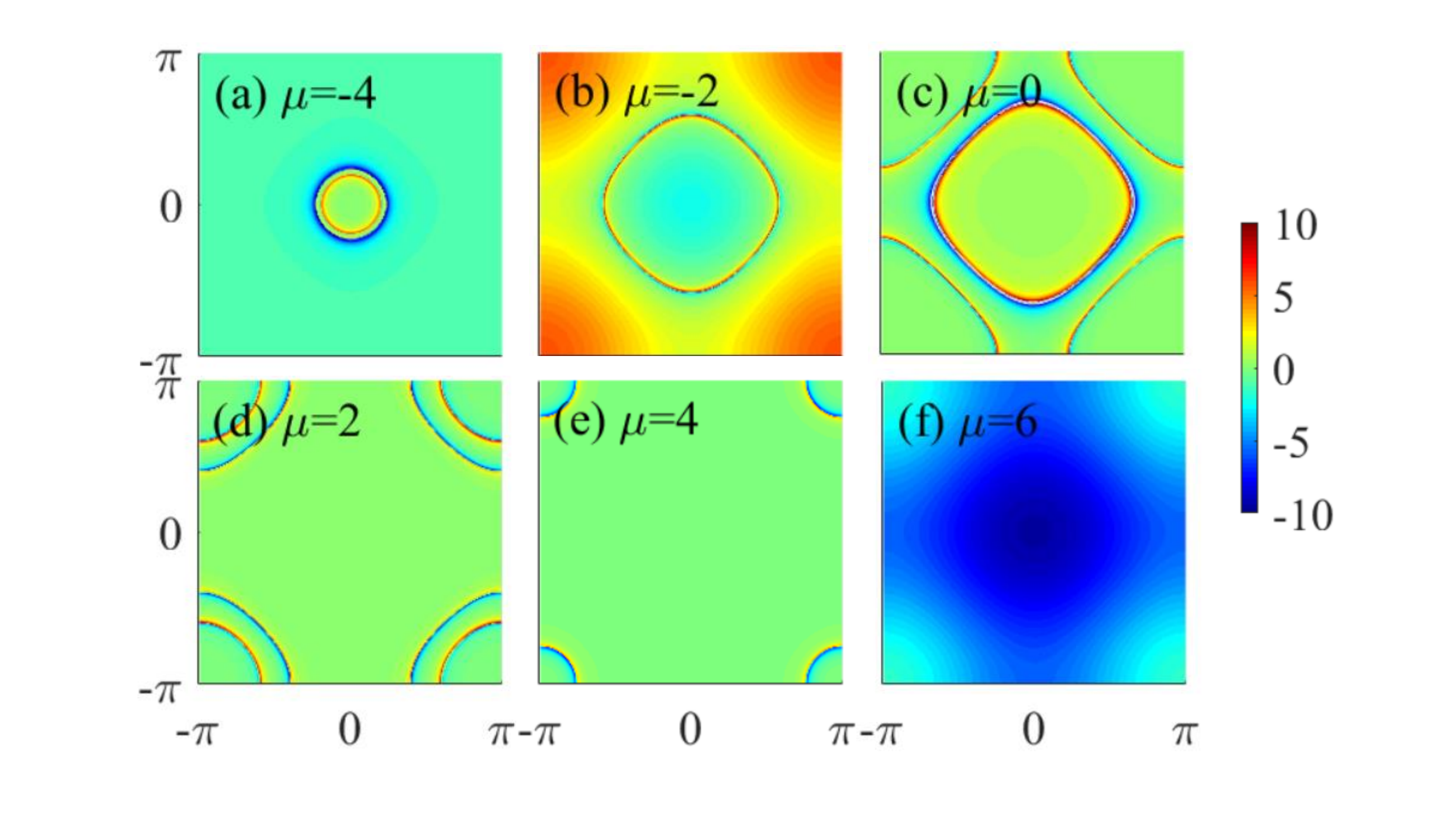}
\caption{\label{fig:11} The real part $f$-electron's zero-frequency Green's function $\mathrm{Re}G_{f}(k_{x},k_{y},\omega=0)$ on $2D$ square lattice with $E_{f}=-2,U=8$. (a) $\mu=-4$; (b) $\mu=-2$; (c) $\mu=0$; (d) $\mu=2$; (e) $\mu=4$; (f) $\mu=6$.}
\end{figure}

In addition, we have shown the $f$-electron's zero-frequency spectral function $A_{f}(k_{x},k_{y},\omega=0)$ on the $2D$ square lattice  in Fig.~\ref{fig:10}, which is able to encode the structure of Fermi surface. Obviously, one observes that tuning chemical potential $\mu$ drives the transition of Fermi surface, which is the counterpart of Lifshitz transition in $2D$.
The corresponding real part of $f$-electron's Green's function at $\omega=0$ has been shown in Fig.~\ref{fig:11}, where clear jump from $-\infty$ to $\infty$ appears indicates the existence of Fermi surface.
\subsection{Relation between particle density and metallic state}
We have seen that in the ground-state of Eq.~\ref{eq4}, the particle density of insulating states is integer ($n=0,2,3,4$) while the metallic CM is generally not an integer. Here, following the treatment in Ref.~\cite{Yamanaka1997} which utilizes the classic proof of  Lieb-Schultz-Mattis (LSM) theorem, we provide an intuitive argument on this point.

At first, we rewrite Eq.~\ref{eq4} into its real space version, whose Hamiltonian reads
\begin{eqnarray}
\hat{H}&=&-t\sum_{j\sigma}(\hat{c}_{j\sigma}^{\dag}\hat{c}_{j+1\sigma}+\hat{c}_{j+1\sigma}^{\dag}\hat{c}_{j\sigma})-\mu\sum_{j\sigma}\hat{c}_{j\sigma}^{\dag}\hat{c}_{j\sigma}\nonumber\\
&+&\sum_{j\sigma}(E_{f}-\mu)\hat{f}_{j\sigma}^{\dag}\hat{f}_{j\sigma}+V\sum_{j\sigma}(\hat{c}_{j\sigma}^{\dag}\hat{f}_{j\sigma}+\hat{f}_{j\sigma}^{\dag}\hat{c}_{j\sigma})\nonumber\\
&+&\frac{U}{N_{s}}
\sum_{j_{1}j_{2}j_{3}j_{4}}\delta_{j_{1}+j_{3}=j_{2}+j_{4}}\hat{f}_{j_{1}\uparrow}^{\dag}\hat{f}_{j_{2}\uparrow}\hat{f}_{j_{3}\downarrow}^{\dag}\hat{f}_{j_{4}\downarrow}.
\end{eqnarray}
Then, consider periodic boundary condition and define the twist operator
\begin{equation}
\hat{U}=e^{i\sum_{j=1}^{N_{s}}\frac{2\pi j}{N_{s}}\sum_{\sigma}(\hat{c}_{j\sigma}^{\dag}\hat{c}_{j\sigma}+\hat{f}_{j\sigma}^{\dag}\hat{f}_{j\sigma})}.
\end{equation}
If we denote the ground-state as $|\Psi_{0}\rangle$, then a new state is constructed by applying $\hat{U}$, i.e. the twisted state $\hat{U}|\Psi_{0}\rangle$. So, one can calculate the energy difference
\begin{eqnarray}
\Delta E&=&\langle\Psi_{0}|\hat{U}^{-1}\hat{H}\hat{U}|\Psi_{0}\rangle-\langle \Psi_{0}|\hat{H}|\Psi_{0}\rangle\nonumber\\
&=&\sum_{\sigma}\sum_{j=1}^{N_{s}}(2-e^{-i2\pi/N_{s}}-e^{i2\pi/N_{s}})t\langle \hat{c}_{j\sigma}^{\dag}\hat{c}_{j+1\sigma}\rangle.\nonumber
\end{eqnarray}
When $N_{s}>>1$, $\Delta E\sim \mathcal{O}(1/N_{s})$, thus there exists at least one low-energy state near ground-state. Furthermore, for the translation operator $\hat{T}$, we have $\hat{T}\hat{U}\hat{T}^{-1}=\hat{U}e^{-i2\pi \hat{n}}$. ($\hat{n}=\frac{1}{N_{s}}\sum_{j\sigma}(\hat{c}_{j\sigma}^{\dag}\hat{c}_{j\sigma}+\hat{f}_{j\sigma}^{\dag}\hat{f}_{j\sigma})$) Assume the ground-state $|\Psi_{0}\rangle$ has particle density $n$ and momentum $P_{0}$, then
\begin{equation}
\hat{T}\hat{U}|\Psi_{0}\rangle=\hat{U}\hat{T}e^{-i2\pi \hat{n}}|\Psi_{0}\rangle=e^{-i2\pi n}e^{-iP_{0}}\hat{U}|\Psi_{0}\rangle,
\end{equation}
which means the twisted state is the eigenstate of momentum $2\pi n+P_{0}$. If $n$ is not an integer, $\hat{U}|\Psi_{0}\rangle$ and $|\Psi_{0}\rangle$ must be orthogonal, thus the system is gapless in this situation and it corresponds to metallic state. In contrast, when $n$ is an integer, we expect $\hat{U}|\Psi_{0}\rangle$ and $|\Psi_{0}\rangle$ are the same state which suggests that there exist no low-energy state and the system should be an insulator. Therefore, we now understand why $n=0,2,3,4$ in our model correspond to insulating states and generic electron's density implies metallic state.
\section{Conclusion and Future direction}\label{sec3}
In conclusion, we have provided a solvable quantum many-body model, whose solvability is due to locality in momentum space and analyzed the ground-state properties in its $1D$ version. Importantly, we find a non-Fermi liquid-like metallic state violating the Luttinger theorem and an interaction-driven Mott insulator in ground-state. The involved quantum phase transition between these two states belongs to the universality of Lifshitz transition. The phase diagram of the $2D$ version of our model is similar to $1D$ and it is expected that the findings in the latter may be generic for all spatial dimensions.

In light of recent theoretical and experimental works on topological Kondo insulator, particularly the unusual quantum oscillation found in SmB$_{6}$ and YbB$_{12}$,\cite{Li2014,Tan2015,Xiang2018,Knoll2015,Zhang2016,Riseborough2017,Baskaran2015,Varma2020,Erten2016,Knolle2017,Rao2019} our solvable model can be modified (with momentum and spin-dependent hybridization strength) to attack topological phases relevant to these interesting $f$-electron compounds, for example a solvable model to give insight into $1D$ topological Kondo insulator is\cite{Zhong2017,Lisandrini,Luo2021}
\begin{eqnarray}
\hat{H}&=&-t\sum_{j\sigma}(\hat{c}^{\dag}_{j\sigma}\hat{c}_{j+1\sigma}+\hat{c}^{\dag}_{j+1\sigma}\hat{c}_{j\sigma})-\mu\sum_{j\sigma}\hat{c}_{j\sigma}^{\dag}\hat{c}_{j\sigma}\nonumber\\
&+&\frac{1}{2}\sum_{j\sigma,\delta=\pm1}V_{j,j+\delta}(\hat{c}^{\dag}_{j+\delta\sigma}\hat{f}_{j\sigma}+\hat{f}_{j\sigma}^{\dag}\hat{c}_{j+\delta\sigma})\nonumber\\
&+&E_{f}\sum_{j\sigma}\hat{f}_{j\sigma}^{\dag}\hat{f}_{j\sigma}+\frac{U}{N_{s}}
\sum_{j_{1}j_{2}j_{3}j_{4}}\delta_{j_{1}+j_{3}=j_{2}+j_{4}}\hat{f}_{j_{1}\uparrow}^{\dag}\hat{f}_{j_{2}\uparrow}\hat{f}_{j_{3}\downarrow}^{\dag}\hat{f}_{j_{4}\downarrow},\nonumber
\end{eqnarray}
with $V_{j,j+1}=V,V_{j,j-1}=-V$. Such exciting possibility will be explored in our future work and we hope our model will be a good starting point to explore unexpected physics in correlated electron systems.

\appendix
\section{Diagonalization of $\hat{H}_{k}$}
Before diagonalizing $\hat{H}_{k}$, we observe that particle number $\hat{n}_{k}=\sum_{\sigma}(\hat{c}_{k\sigma}^{\dag}\hat{c}_{k\sigma}+\hat{f}_{k\sigma}^{\dag}\hat{f}_{k\sigma})$ and spin $\hat{S}^{z}_{k}=\frac{1}{2}(\hat{c}_{k\uparrow}^{\dag}\hat{c}_{k\uparrow}-\hat{c}_{k\downarrow}^{\dag}\hat{c}_{k\downarrow}+\hat{f}_{k\uparrow}^{\dag}\hat{f}_{k\uparrow}-\hat{f}_{k\downarrow}^{\dag}\hat{f}_{k\downarrow})$, are both conserved in $\hat{H}_{k}$. Therefore, the $16\times16$-matrix of $\hat{H}_{k}$ has the block-diagonal form as $H_{k}=1\oplus2\oplus2\oplus1\oplus1\oplus4\oplus2\oplus2\oplus1$. The first $1$ means $|0000\rangle$ with $n_{k}=0,S_{k}^{z}=0$ is the eigen-state with eigen-energy $0$. The second $2$ encodes that $H_{k}=\left(
                                                           \begin{array}{cc}
                                                             \xi_{k}^{c} & V \\
                                                             V & \xi_{k}^{f} \\
                                                           \end{array}
                                                         \right)$
in the subspace formed by $|1000\rangle$ and $|0010\rangle$ with $n_{k}=1,S_{k}^{z}=1/2$. Here, we have introduced $\xi_{k}^{c}=\varepsilon_{k}^{c}-\mu,\xi_{k}^{f}=\varepsilon_{k}^{f}+E_{f}-\mu $ to simplify the expression. The third $3$ gives the same $2\times2$-matrix, which is
formed by $|0100\rangle$ and $|0001\rangle$ with $n_{k}=1,S_{k}^{z}=-1/2$. The fourth and fifth $1$ produce identical energy $\xi_{k}^{c}+\xi_{k}^{f}$ for $|1010\rangle$ state with $n_{k}=2,S^{z}_{k}=1$ and
$|0,1,0,1\rangle$ state with $n_{k}=2,S^{z}_{k}=-1$. Now, we come with the sixth $4$, whose subspace is formed by $|1100\rangle$,$|1001\rangle$,$|0110\rangle$ and $|0011\rangle$ with $n_{k}=2,S_{k}^{z}=0$. Therefore, $\hat{H}_{k}$ has the following $4\times4$-matrix representation,

\begin{equation}
  H_{k}=\left(
    \begin{array}{cccc}
      2\xi_{k} & V & -V & 0 \\
      V & \xi_{k}^{c}+\xi_{k}^{f} & 0 & V \\
      -V & 0 & \xi_{k}^{c}+\xi_{k}^{f} & -V \\
      0 & V & -V & 2\xi_{k}^{f}+U \\
    \end{array}
  \right).
\end{equation}
Although analytic expression of eigen-energy and eigen-state is available, it is too complicated to use, so we diagonalize above matrix numerically.

Then, the seventh and eighth $2$ give the same $H_{k}=\left(
                                                           \begin{array}{cc}
                                                             2\xi_{k}^{c}+\xi_{k}^{f} & -V \\
                                                             -V & \xi_{k}^{c}+2\xi_{k}^{f}+U \\
                                                           \end{array}
                                                         \right)$
, which is formed by $|1110\rangle$,$|1011\rangle$ with $n_{k}=3,S_{k}^{z}=1/2$ and $|1101\rangle$,$|0111\rangle$ with $n_{k}=3,S^{z}_{k}=-1/2$, respectively. The last $1$ means that $|1111\rangle$ state has eigen-energy $2\xi_{k}^{c}+2\xi_{k}^{f}+U$.

Based on above results, we conclude that in terms of Fock states $|n_{1},n_{2},n_{3},n_{4}\rangle$,($n_{i}=0,1$) the Hamiltonian $\hat{H}_{k}$ has been diagonalized and we obtain $16$ eigen-energy $E_{k}(i)$ and eigen-state $\psi_{k}(i)$.($i=1,2...16$) With the same spirit, one can construct matrix expression like fermionic operator $\hat{c}_{k\sigma},\hat{f}_{k\sigma}$, which are useful to calculate Green's function and spectral function.

\section{Phase diagram for $E_{f}=0$ and $E_{f}=2$}
\begin{figure}
\includegraphics[width=1.0\linewidth]{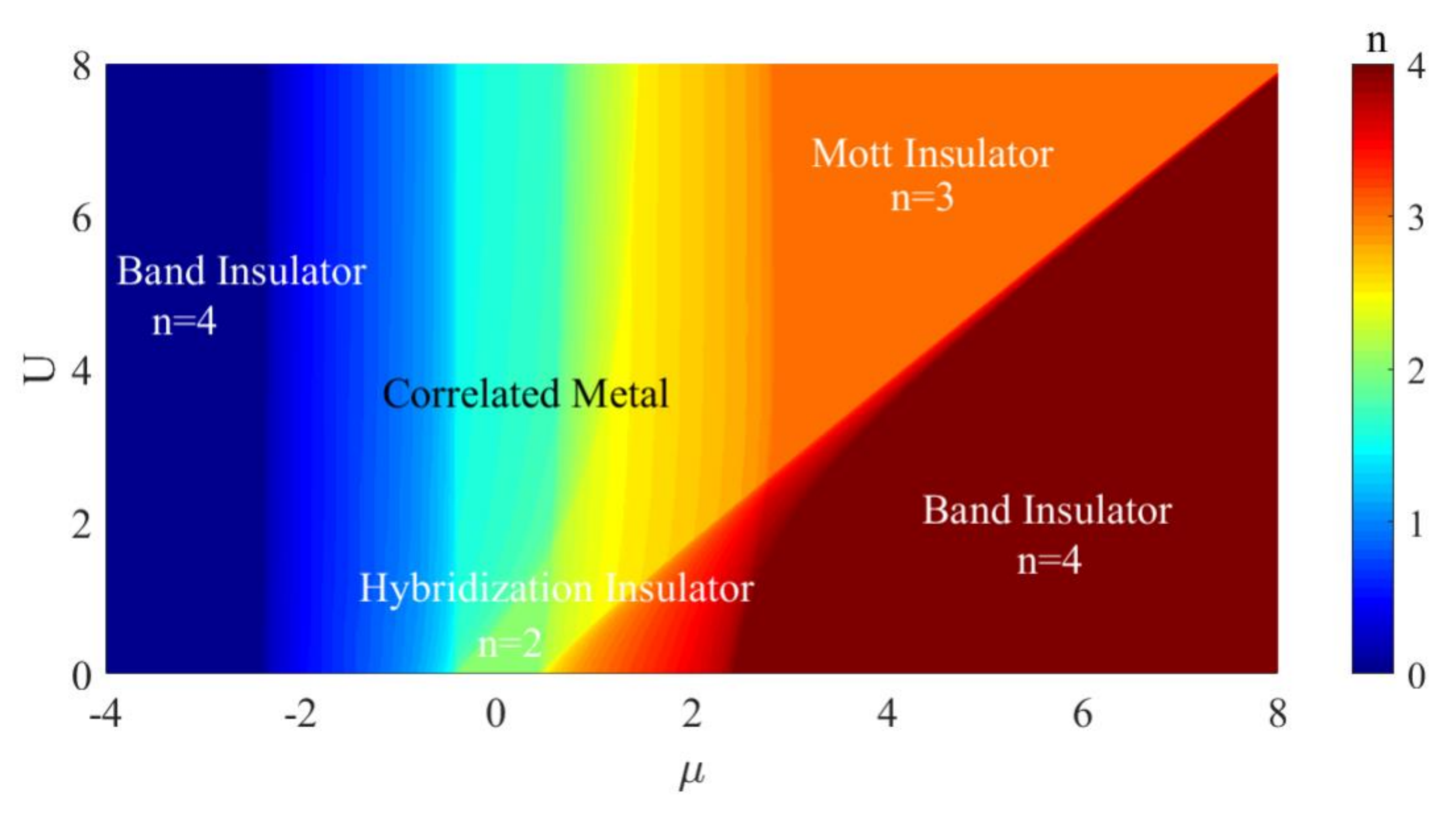}
\caption{\label{fig:B1} Ground-state phase diagram of Eq.~\ref{eq3} with fixed $E_{f}=0$.}
\end{figure}

\begin{figure}
\includegraphics[width=1.0\linewidth]{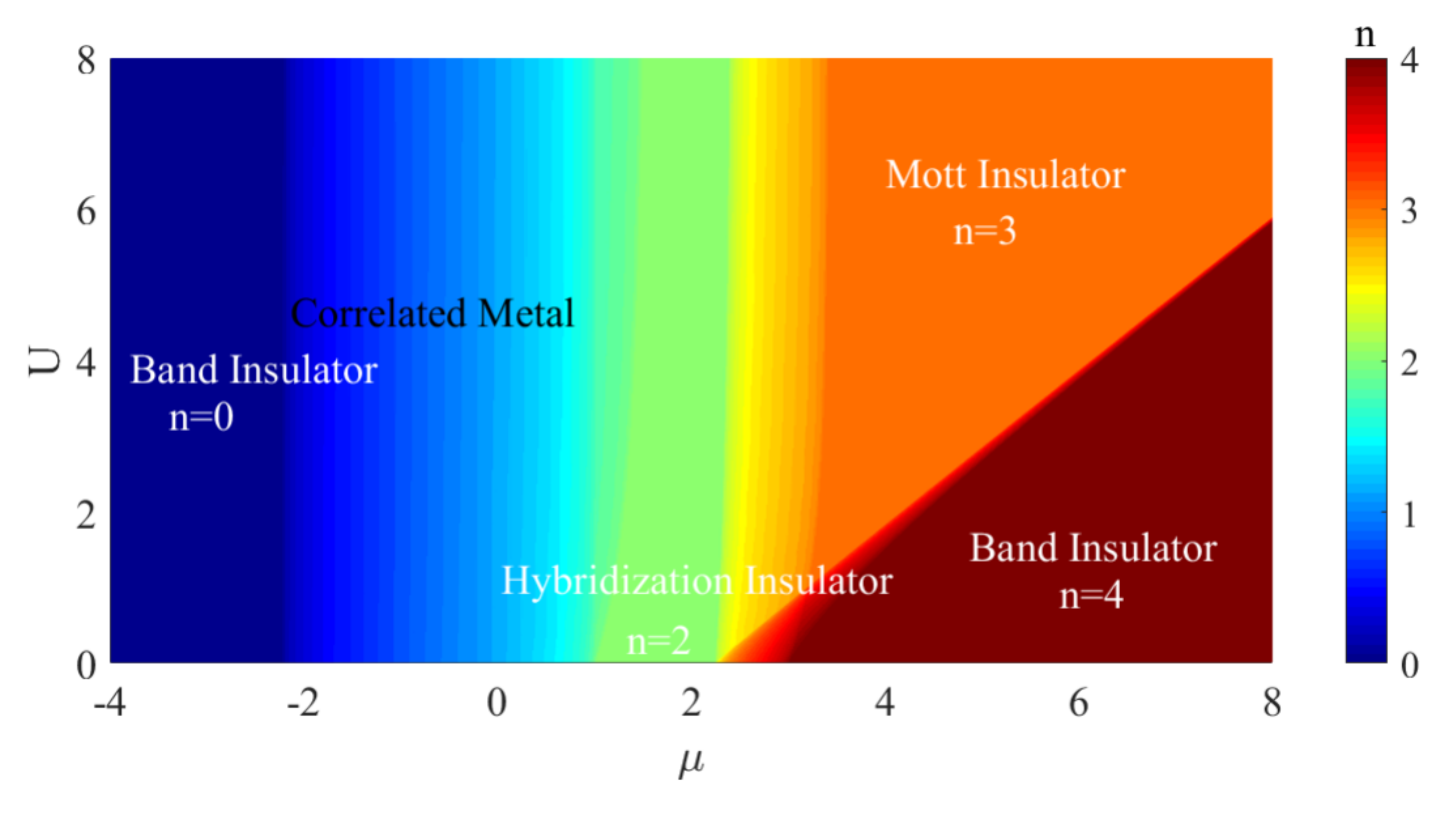}
\caption{\label{fig:B2} Ground-state phase diagram of Eq.~\ref{eq3} with fixed $E_{f}=2$.}
\end{figure}

Here, we plot the ground-state phase diagram for $1D$ model with $E_{f}=0$ and $E_{f}=2$ in Fig.~\ref{fig:B2} and Fig.~\ref{fig:B1}. Other parameters are chosen as the $E_{f}=-2$ case. We see that the structure of all phase diagrams for $E_{f}=-2,0,2$ is similar and the generic physics does not change. Additionally, we observe that increasing $E_{f}$ has the effect to enlarge the $n=2$ HI and $n=3$ MI regime.

\section{Finite temperature}
\begin{figure}
\includegraphics[width=1.1\linewidth]{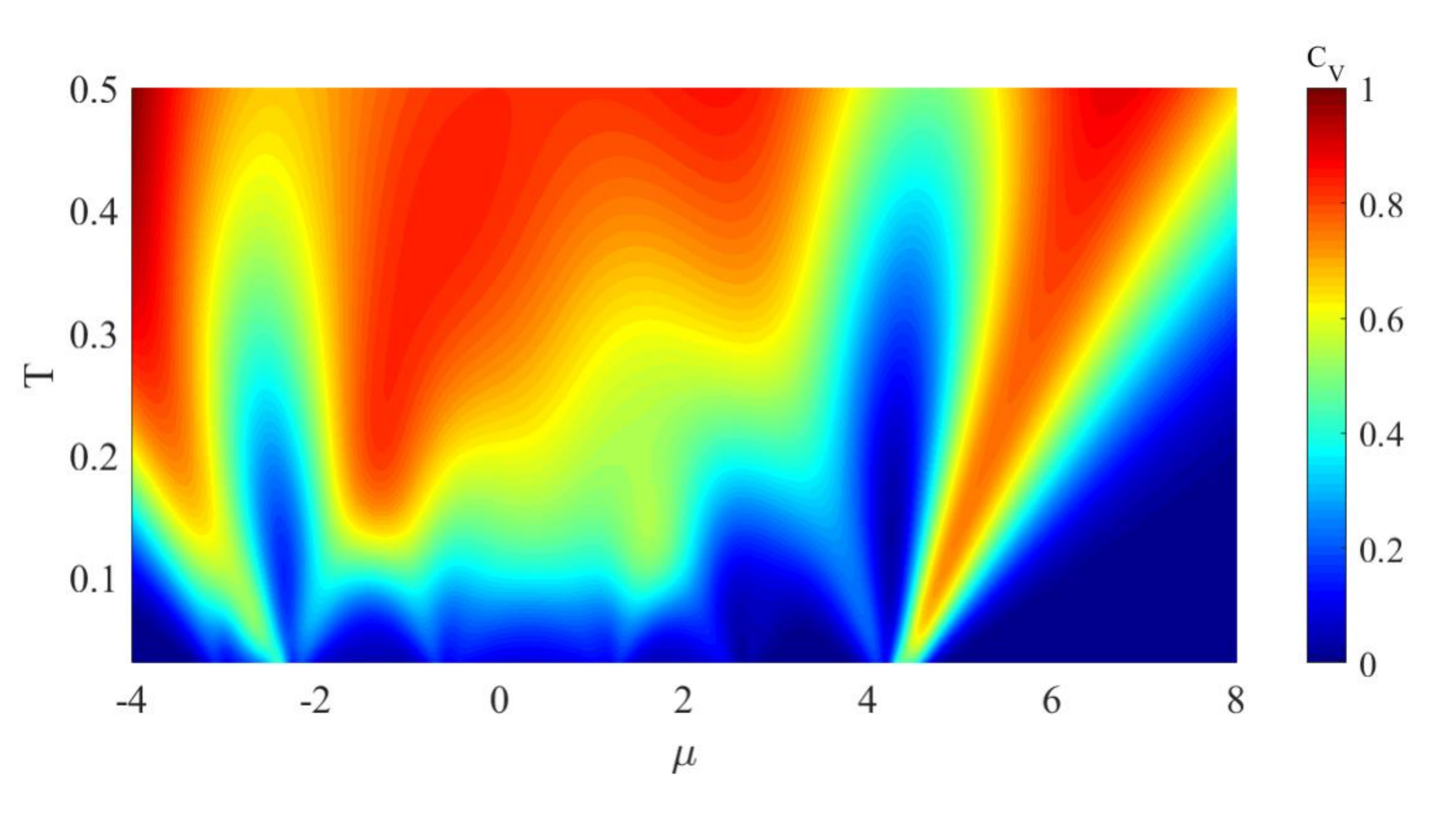}
\caption{\label{fig:B3} The heat capacity $C_{V}$ versus $\mu$ and $T$ with fixed $E_{f}=-2,U=6$.}
\end{figure}

After presenting results on ground-state in the main text, we now study the finite temperature case. At finite-$T$, the thermodynamics of our model is determined by
its free energy density $f$, which is related to partition function $\mathcal{Z}$ as
\begin{eqnarray}
&&f=-\frac{T}{N_{s}}\ln\mathcal{Z},\nonumber\\
&&\mathcal{Z}=\mathrm{Tr }e^{-\beta \hat{H}}=\prod_{k}\mathrm{Tr} e^{-\beta \hat{H}_{k}}=\prod_{k}\left(\sum_{j=1}^{16}e^{-\beta E_{k}(j)}\right).
\end{eqnarray}
Here, one notes that the partition function is easy to calculate since each $k$-state contributes independently. Then, the typical thermodynamic quantity, i.e. the heat capacity, is calculated by standard thermodynamic relation $C_{V}=-T\frac{\partial^{2}f}{\partial T^{2}}$, which has been shown in Fig.~\ref{fig:B3}.

In addition to $C_{V}$, one can also calculate spin susceptibility $\chi_{s}$ if inserting Zeeman energy term $\hat{H}_{h}=-h\hat{S}_{k}^{z}=-\frac{h}{2}(\hat{c}_{k\uparrow}^{\dag}\hat{c}_{k\uparrow}-\hat{c}_{k\downarrow}^{\dag}\hat{c}_{k\downarrow}+\hat{f}_{k\uparrow}^{\dag}\hat{f}_{k\uparrow}-\hat{f}_{k\downarrow}^{\dag}\hat{f}_{k\downarrow})$ into Hamiltonian $\hat{H}_{k}$. Then, it follows that the magnetization $M=-\frac{\partial f}{\partial h}$ and $\chi_{s}=\frac{\partial M}{\partial h}=-\frac{\partial^{2}f}{\partial h^{2}}$.

\end{document}